\documentclass[nofootinbib,amsmath,twocolumn,notitlepage,preprintnumbers]{revtex4-1}
\usepackage{multirow}
\usepackage{amssymb, esvect, amsmath, graphicx, latexsym, amsthm, slashed, eso-pic}
\usepackage{xcolor}
\usepackage{hyperref}
\usepackage{amsmath}

\newcommand{\beq}{\begin{equation}} \newcommand{\eeq}{\end{equation}}
\newcommand{\bea}{\begin{eqnarray}} \newcommand{\eea}{\end{eqnarray}}

\def\lsim{\mathrel{\raise.3ex\hbox{$<$\kern-.75em\lower1ex\hbox{$\sim$}}}}
\def\gsim{\mathrel{\raise.3ex\hbox{$>$\kern-.75em\lower1ex\hbox{$\sim$}}}}

\newcommand{\Eq}[1]{Eq.~(\ref{#1})}

\newcommand{\be}{\begin{eqnarray}}
\newcommand{\ee}{\end{eqnarray}}

\newcommand{\benum}{\begin{enumerate}}
\newcommand{\eenum}{\end{enumerate}}
\newcommand{\bi}{\begin{itemize}}
\newcommand{\ei}{\end{itemize}}

\newcommand{\mpl}{M_{\rm Pl}}

\newcommand{\brac}[2]{ \left( \frac{#1}{#2} \right) }

\begin{document}

\preprint{FERMILAB-PUB-22-432-T
}

\title{Cosmological Magnetic Fields from Primordial Kerr-Newman Black Holes}

\author{Dan Hooper$^{a,b,c}$}
\thanks{ORCID: http://orcid.org/0000-0001-8837-4127}

\author{Aurora Ireland$^{c,d}$}
\thanks{ORCID: https://orcid.org/0000-0001-5393-0971}

\author{Gordan Krnjaic$^{a,b,c}$}
\thanks{ORCID: http://orcid.org/0000-0001-7420-9577}

\affiliation{$^a$University of Chicago, Kavli Institute for Cosmological Physics, Chicago IL, USA}
\affiliation{$^b$University of Chicago, Department of Astronomy and Astrophysics, Chicago IL, USA}
\affiliation{$^c$Fermi National Accelerator Laboratory, Theoretical Astrophysics Group, Batavia, IL, USA}
\affiliation{$^d$University of Chicago, Department of Physics, Chicago IL, USA}

\date{\today}

\begin{abstract}

The origin of our universe's cosmological magnetic fields remains a mystery. In this study, we consider whether these magnetic fields could have been generated in the early universe by a population of charged, spinning primordial black holes. To this end, we calculate the strength and correlation length of the magnetic fields generated by this population, and describe their evolution up to the current epoch. We find that near-extremal black holes in the mass range $ M \sim 10^{28} -10^{36} \, {\rm g}$ could potentially generate magnetic fields with present day values as large as $B \sim 10^{-20} - 10^{-15} \, {\rm G}$; those with $M \gtrsim10^{38} \, \rm{g}$ could have produced even larger fields $B \gtrsim 10^{-14} \, {\rm G}$. To motivate this scenario, we briefly discuss how new physics may have induced a chemical potential which could have briefly maintained the black holes in an electrically charged state in the early universe. Finally, we comment on a correlation between the parameters of the cosmological magnetic field and the stochastic gravitational wave background coming from the merger of primordial black hole binaries as the primary observable signature of this scenario.

\end{abstract}

\maketitle

\section{Introduction}

According to the standard paradigm\footnote{Certain inflationary mechanisms are capable of producing magnetic fields of sufficient strength ($\gtrsim 0.1$ nG on Mpc scales) that they can be adiabatically compressed to explain the $\mu$G strength fields in galaxies today without invoking the dynamo mechanism~\cite{Mandal:2022,Pogosian:2019}.}, the magnetic fields present within galaxies and galaxy clusters were generated through the amplification of  preexisting, but much weaker, magnetic fields through the dynamo mechanism~\cite{Parker:1955zz,Kulsrud:1999bg,Brandenburg:2004jv,Kulsrud:2007an,Durrer:2013pga}. This process is effective, however, only if a non-zero magnetic field is present for the dynamos to amplify. The origin of these magnetic field ``seeds," which were present at the onset of structure formation, remains an open question and has generated a great deal of speculation~\cite{Kronberg:1993vk,Grasso:2000wj,Widrow:2002ud,Kulsrud:2007an,Kandus:2010nw,Widrow:2011hs}. It has been proposed that primordial magnetic fields could arise within the context of inflation~\cite{Turner:1987bw,Ratra:1991bn,Martin:2007ue,Subramanian:2009fu,Kandus:2010nw,Motta:2012rn,Jain:2012ga} or during phase transitions that took place in the early universe~\cite{Vachaspati:1991nm,Enqvist:1993np,Grasso:1997nx,Joyce:1997uy,Caprini:2009yp,Kamionkowski:1993fg,Huber:2008hg}. None of these scenarios is completely satisfactory, though, and each faces its own challenges. In particular, it is difficult for these mechanisms to produce fields of sufficiently large correlation length so as to survive until today.

The origin of the primordial magnetic field is somewhat obscured by the complicated plasma and magnetohydrodynamics processes that have taken place over cosmic time. One can attempt, however, to constrain the properties of the seed field by studying the magnetic fields found within the voids of the intergalactic medium, where primordial fields could exist in a relatively pristine state. In such environments, the evolution of the magnetic field would be largely driven by the expansion of the universe, leading to the dilution of the field strength as $B \propto a^{-2}$ (corresponding to $\rho_B \propto a^{-4}$), and to the growth of the field's correlation length as $\xi \propto a$.

In this article, we consider the possibility that primordial magnetic fields may have been generated in the early universe by a subdominant population of primordial black holes. In order to produce a non-zero magnetic field, these black holes must have been both spinning and electrically charged, corresponding to the Kerr-Newman solution. In our scenario, this population is temporarily charged in the early universe due to a nonzero chemical potential, which eventually relaxes to zero, at which point the black holes discharge. Afterwards, the magnetic fields evolve according to Hubble expansion and the (now neutral) black holes constitute a present day dark matter abundance. While such a scenario is admittedly quite speculative and involves some rather exotic elements, we find that astrophysically interesting magnetic fields could have potentially been generated by such objects. 


\section{Kerr-Newman Black Holes}
\label{sec:KNBH}

Generating a magnetic field requires both an electromagnetic current and a departure from spherical symmetry. For this reason, we are interested here in black holes that are both charged and rotating. Such Kerr-Newman black holes are entirely characterized by their mass, $M$, angular momentum, $J$, and charge, $Q$. In Boyer-Lindquist coordinates, the geometry associated with such an object is described by the following line element~\cite{Kerr:1963,Newman:1965,Misner:1973}:
\be
\label{KNmetric}
	ds^2 &=& - \frac{\Delta}{\rho^2} (dt - \alpha \sin^2 \theta \, d\phi)^2 + \frac{\rho^2}{\Delta} dr^2 \\
	&&+ \rho^2 d\theta^2 + \frac{\sin^2 \theta}{\rho^2} \left[ (r^2+\alpha^2) d\phi - \alpha dt \right]^2, \nonumber
\ee
where $\alpha = J/M$, and we have defined
\be\label{rhoeq}
\rho^2 = r^2 + \alpha^2 \cos^2 \theta, ~~
\Delta = r^2 + \alpha^2 - \frac{2 M r}{\mpl^2}  + \frac{Q^2}{\mpl^2},~~~~
\ee
and $\mpl = 1.22 \times 10^{19}$ GeV is the Planck mass.
The charge and angular momentum of a black hole are constrained to lie within the following domain:
\be
	\alpha^2\mpl^2 + Q^2 \le \frac{M^2}{\mpl^2} \,.
\ee
From the metric, we see that the Kerr-Newman black hole has two horizons located at
\begin{equation}
	r_{\pm} = \frac{1}{\mpl^2} \left( M \pm \sqrt{M^2 - \alpha^2 \mpl^4 - Q^2 \mpl^2} \right).
\end{equation}
Integrating over the angular volume element evaluated on the $r = r_+$ hypersurface yields the area of the event horizon
\begin{equation}\label{KNarea}
	A = 4\pi \, (r_+^2 + \alpha^2).
\end{equation}
From the Killing vector associated with the event horizon, the surface gravity can be written as~\cite{Misner:1973}
\begin{equation}
	\kappa = \frac{2\pi}{A} (r_+ - r_-).
\end{equation}
These two quantities are related to a black hole's temperature and entropy as follows~\cite{Bardeen:1973}:
\begin{eqnarray}
\label{gentemp}
	T_{\rm BH} &=& \frac{\kappa}{2\pi}  =  \frac{r_+ - r_-}{4\pi (r_+^2 + \alpha^2)} \\
	S_{\rm BH} &=& \frac{A}{4}  =  \pi (r_+^2 + \alpha^2). 
\end{eqnarray}
These expressions, in conjunction with the fact that the mass of a black hole can be identified with energy, yields the first law of black hole thermodynamics:
\begin{equation}\label{first}
	dM = \frac{\mpl^2}{8\pi} \,\kappa \, dA + \Omega \, dJ + \Phi \, dQ,
\end{equation}
where $\Omega$ and $\Phi$ are the angular velocity and the electrostatic potential of the black hole. Note that the quantities $\kappa$ (and hence $T_{\rm BH}$), $\Omega$, and $\Phi$ are constant over the horizon. In order to obtain explicit forms for $\Omega$ and $\Phi$ in the context of a Kerr-Newman black hole, we need to take the differential of the area given in Eq.~(\ref{KNarea}). After some algebra, we can write
\begin{equation}
	\frac{\mpl^2}{8\pi} \,\kappa \, dA = \frac{r_+^2 dM}{r_+^2 + \alpha^2}  -\frac{r_+ Q dQ}{r_+^2 + \alpha^2} -  \frac{M \alpha  d\alpha}{r_+^2 + \alpha^2}.
\end{equation}
Substituting $-\alpha M d\alpha = -\alpha dJ + \alpha^2 dM$ and inserting the explicit form for $\kappa$, we arrive at the following expression:
\begin{equation}
	dM = \frac{\mpl^2}{4} \, T_{\rm BH}\, dA +  \frac{\alpha dJ}{r_+^2 + \alpha^2}  + \frac{r_+ Q dQ  }{r_+^2 + \alpha^2}.
\end{equation}
Comparing this to Eq.~(\ref{first}), we can determine the black hole's angular velocity and electrostatic
potential: 
\be
\label{apotential}
	\Omega = \frac{\alpha}{r_+^2 + \alpha^2}, ~~~	\Phi = \frac{r_+ Q}{r_+^2 + \alpha^2}.
\ee

\section{Generating Cosmological Magnetic Fields}
\label{sec:B}


We begin by considering an isolated black hole whose mass, angular momentum, and charge are not appreciably evolving with time, hence neglecting the possible effects of Hawking evaporation
 and accretion. This stationary geometry is described by the Kerr-Newman metric given in Eq.~(\ref{KNmetric}). Technically this is just one half of the complete solution to the coupled Einstein-Maxwell equations, which describe the interplay between the dynamical metric and electromagnetic field. For a full solution, we must also specify the vector potential $A_\mu$~\cite{Misner:1973} 
\be
\label{EMpotential}
	A_\mu dx^\mu  = - \frac{Q r}{r^2 + \alpha^2 \cos^2 \theta} \left(   dt - \alpha \sin^2 \theta \, d\phi   \right).
\ee
Using the field strength, $F_{\mu \nu} = \nabla_\mu A_\nu - \nabla_\nu A_\mu$, the $E$ and $B$ fields are given by
%
\be\label{eq:fields}
	\vec{E} &=&  \frac{Q (r^2 - \alpha^2 \cos^2 \theta)}{\rho^4} \, \hat{r} - \frac{2 Q \alpha^2 \cos \theta \sin \theta}{\rho^4}  \, \hat{\theta}, \\
	\vec{B} &=& \frac{  Q \alpha }{r }  \left[   \frac{2(\alpha^2 + r^2) \cos \theta}{\rho^4}  \, \hat{r}  + \frac{ (r^2 - \alpha^2 \cos^2 \theta) \sin \theta}{\rho^4}  \, \hat{\theta} \right]. \nonumber
\ee
Note that the $E_\phi$ and $B_\phi$ components are both vanishing, since we've taken the black hole to be rotating in the $\hat{\phi}$ direction. Also note that in the $r \rightarrow \infty$ limit, these fields have the expected asymptotic forms:
\be
	\lim_{r \rightarrow \infty} \vec{E} &=& \frac{Q}{r^2} \, \hat{r} + \mathcal{O}\left(\frac{1}{r^3} \right) , \\
	\lim_{r \rightarrow \infty} \vec{B} &=& \frac{ Q        \alpha      }{r^3} \left(   2 \cos \theta    \,   \hat{r} +  \sin \theta \, \hat{\theta}\right) + \mathcal{O}\left(\frac{1}{r^4} \right). \nonumber 
\ee
%

In considering the case of an isotropic population of black holes,\footnote{A possible objection to this scenario is that the black holes might act as an ensemble of magnetic dipoles which interact to form domains of some characteristic scale. This will not be applicable in this case, however, as we will consider black hole number densities which are sufficiently small such that no more than one black hole will be present in a given Hubble radius at early times.} it will be useful to have an expression for the magnetic field of a single black hole averaged over a sphere of radius $R > r_+$. We adopt the volume-averaged convention\footnote{We have confirmed numerically that our definition coincides with the RMS average value, $B_{\text{RMS}}^2 = \frac{1}{V} \int d^3x \, \vec{B}^2$, up to an $\mathcal{O}(1)$ factor.}:
\begin{equation}
	\langle \vec{B} \rangle = \frac{1}{V} \int_V d^3x \, \vec{B} \,,
\end{equation}
where $V = 4\pi R^3/3$ is the volume of the sphere over which we are averaging. Starting from Eq.~(\ref{eq:fields}) and omitting the algebraic details, the volume-averaged magnetic field magnitude can be written as
\begin{equation}\label{Bav}
	\langle  B  \rangle = \frac{3Q}{R^2} \left[ \left( 1+ \frac{R^2}{\alpha^2} \right) \tan^{-1} \left( \frac{\alpha}{R} \right) - \frac{R}{\alpha} \right].
\end{equation}
In the $\alpha \ll R$ limit, the average magnetic field reduces to $\langle B \rangle \approx 2Q \alpha /R^3$. This limit will be applicable throughout our entire parameter space of interest.


The primordial magnetic field is also characterized by a correlation length $\xi$, which
governs the extent to which diffusion and damping will suppress any magnetic fields that are generated by black holes in the early universe. On scales greater than the magnetic diffusion length $\ell_{\rm diff}$, diffusive effects can be neglected, so the comoving field is said to be ``frozen in" and $\xi$  grows linearly with the scale factor of the universe. 

Formally $\xi_B$ is defined as the length scale after which correlations, as quantified by the two-point correlation function, fall off exponentially. Intuitively we expect this to coincide approximately with the average distance between neighboring primordial black holes, as it is at this length scale that magnetic field lines will begin to interfere with and wash out one another:
\begin{equation}
\label{rav}
\xi  \sim  \brac{3}{4\pi n_{\rm BH}}^{1/3} =  \left( \frac{45}{2\pi^3 g_{\star}(T)} \frac{M}{f_{\text{BH}} T^4} \right)^{1/3},
\end{equation}
where $f_{\rm BH} = M n_{\rm BH} /\rho_R(T)$ is the energy fraction in black holes relative to that in radiation at the time of magnetogenesis and $g_\star(T)$ is the number of effective relativistic degrees of freedom at temperature $T$. 

Once a magnetic field is generated at some initial temperature, $T_i$, there are several processes which can affect its evolution, including small scale damping, diffusion, and the expansion of the universe \cite{Durrer:2013pga,Grasso:2000wj}. We will make the simplifying assumption that the initial correlation length is sufficiently large that we do not need to account for the former effects $\xi_i > \ell_{\rm diff}$, and focus solely on the impact of Hubble expansion. We will later verify that this assumption is self-consistent for all parameter space of interest. In an expanding universe, the magnetic field redshifts as $B \propto a^{-2}$, while the correlation length grows as $\xi \propto a$. These scalings are manifest when writing $B$ and $\xi$ in terms of temperature: 
\be
\label{BT}
B(T) &=& B_i \left( \frac{T}{T_i} \right)^2  \left[\frac{g_{\star, S}(T)}{g_{\star, S}(T_i)}\right]^{2/3} 
\\ 
\label{xiT}
 \xi(T) &=& \xi_i \left( \frac{T_i}{T} \right)  \left[\frac{g_{\star, S}(T_i)}{g_{\star, S}(T)}\right]^{1/3}, 
\ee
where $g_{\star, S}(T)$ is the effective number of degrees-of-freedom in entropy,
and the initial values at magnetogenesis, $B_i$ and $\xi_i$, can be related 
to black hole parameters using Eqs.~(\ref{Bav}) and~(\ref{rav}), with $R = \xi_i$. 
Defining the following dimensionless parameters:
\begin{equation}
	\alpha_\star \equiv \alpha \frac{\mpl^2}{M} = J \frac{\mpl^2}{M^2}, ~~~~ Q_\star \equiv Q \frac{\mpl}{M},
\end{equation}
the magnetic field from Eq.~(\ref{BT}) can be written as
\be
\label{eq:B}
	\langle B_0  \rangle
	&=&   
	\frac{4 \pi^3   \alpha_\star Q_\star f_{{\rm BH}, i}   g_\star(T_i)  T_i^2  T_{0}^2 M}{45  \mpl^3}    
	 \, \left[\frac{g_{\star, S}(T_{0})}{g_{\star, S}(T_i)}\right]^{2/3} \!\!\! ,~~~~~
\ee
where present day values are denoted by a ``$0$" subscript, $T_0 = 2.725$ K is the CMB temperature,
 and $f_{{\rm BH},i}$ is the black hole energy fraction at $T_i$. In order to express this in terms of current observables, we apply the conservation of entropy:
\be
\label{eq:gstar-ratio}
 \frac{g_{\star, S}(T_0)}{g_{\star, S}(T_{i})}  = \brac{a_i T_i}{a_0 T_0}^3,
\ee
where $a_{i,0}$ is the scale factor at the corresponding epoch. 
Noting also that the initial black hole energy density at magnetogenesis satisfies 
\be
\label{eq:rhos}
\rho_{\rm BH}(T_i)  = f_{{\rm BH},i} \brac{\pi^2 g_\star(T_i)  T_i^4}{30} = \Omega_{\rm BH} \rho_c \brac{a_0}{a_i}^3,~~~~
\ee
where $\Omega_{\rm BH}  \equiv \rho_{\rm BH}/\rho_c$ is the present day energy density in black holes relative to the critical density $\rho_c \approx 4 \times 10^{-47}$ GeV$^4$, we can rewrite \Eq{eq:B} as
 \begin{eqnarray}
 \label{eq:altB}
	\langle B_0  \rangle &=& \frac{8 \pi  \alpha_{\star} Q_{\star}  \Omega_{\rm BH} \rho_{c}  M}{3 M^3_{\rm Pl} } \frac{T_i}{T_0} \left[\frac{g_{\star, S}(T_i)}{g_{\star, S}(T_{0})}\right]^{1/3}	 \\
	&\approx& 6 \times 10^{-16} \, {\rm G} \,\,  \bigg(\frac{Q_\star \alpha_\star}{0.5}\bigg) \left( \frac{\Omega_{{\rm BH}}}{0.01} \right) \brac{M}{M_\odot} 
 \left( \frac{T_i}{\text{GeV}} \right),\nonumber
 \end{eqnarray}
 where in the last line we have used $g_\star = g_{\star, S}  = 73$ at \mbox{$T_i = 1$ GeV}. 
Note that in terms of $\alpha_\star$ and $Q_\star$, the extremality condition is $\alpha_\star^2 + Q_\star^2 \leq 1$, which implies $\alpha_{\star} Q_{\star} \leq 0.5$.
Similarly, combining Eqs.~(\ref{rav}) and (\ref{xiT}), the present day correlation length can be written as
\be
	\xi_0 &=& \frac{1 }{T_{0}} \brac{45 M }{2 \pi^3 g_\star(T_i)  \, f_{{\rm BH}, i}  \,T_i}^{1/3}     \left[  \frac{g_{\star, S}(T_i)}{g_{\star, S}(T_{0})}\right]^{1/3}\!\!.
\ee
Using Eqs.~(\ref{eq:gstar-ratio}) and~(\ref{eq:rhos}), we obtain 
\be
\xi_0 =  \left( \frac{3 M}{4 \pi \, \Omega_{\rm BH} \rho_{c}} \right)^{1/3}  \!\!\! \approx
0.6 \, {\rm kpc}  \, \,  \left( \frac{0.01}{\Omega_{\rm BH}} \right)^{1/3} \!\! \left( \frac{M}{M_\odot} \right)^{1/3}.~~~~~~~ 
\ee
Naively applying Eq.~\ref{eq:altB}, it might appear that arbitrarily strong magnetic fields could be generated by black holes at sufficiently high temperatures, $T_i \gg {\rm GeV}$. Black holes of a given mass, however, can only be formed once $M> M_H$, where $M_H$ is the mass contained within the horizon:
\be
\label{MH}
M_H = \frac{M_{\rm Pl}^2}{2 H_i} \approx  0.06 M_\odot \,\, \bigg(\frac{{\rm GeV}}{T_i}\bigg)^2 \, \bigg(\frac{73}{g_{\star}(T_i)}\bigg)^{1/2}.~~~
\ee
By evaluating Eq.~(\ref{eq:altB}) at $M_H$, we find the following upper limit for the magnetic field strength that could be generated by spinning, charged black holes:
\begin{eqnarray}
\label{Bmax1}
\langle B_0 \rangle_{\rm max} &\approx& 4 \times 10^{-17} \, {\rm G} \, \,  \left( \frac{Q_\star \alpha_\star}{0.5}\right) \left( \frac{\Omega_{{\rm BH}}}{0.01} \right) 
 \left( \frac{\rm GeV}{T_i} \right).~~~~~~~~~
\end{eqnarray}
Alternatively, in terms of the horizon mass, this maximum magnetic field 
can be written as
\be
\label{Bmax2}
\langle B_0 \rangle_{\rm max}  \approx 1.5 \times 10^{-16} \, {\rm G}  \, \,  \left( \frac{Q_\star \alpha_\star}{0.5} \right) \left( \frac{\Omega_{{\rm BH}}}{0.01} \right)  \left( \frac{M_H}{ M_\odot} \right)^{1/2}.~~~~~~
\ee

\section{Potentially Viable Parameter Space}
\label{sec:parameterspace}

\begin{figure}[t]
\hspace{-1cm}
\includegraphics[width=0.485\textwidth]{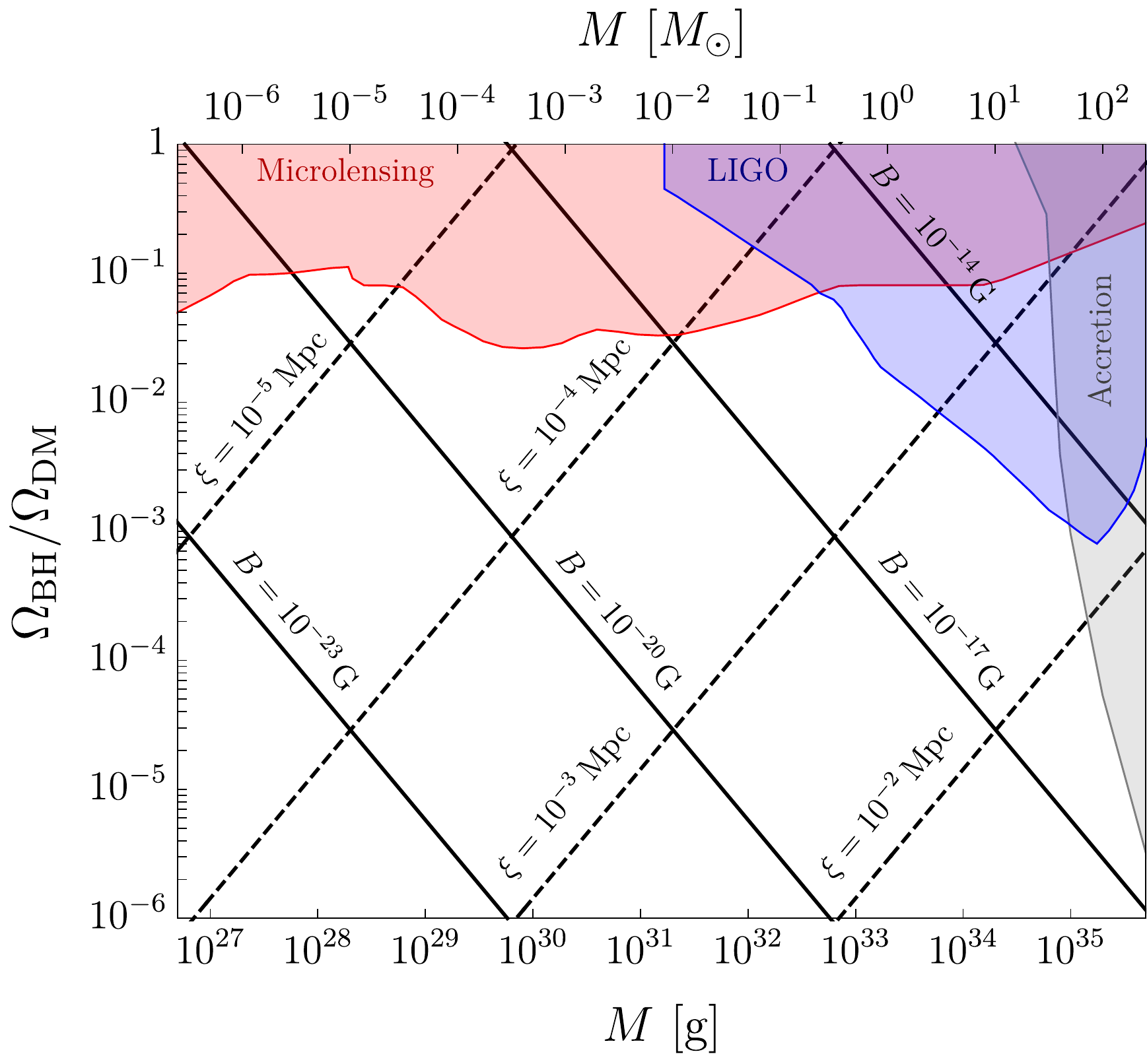}
\caption{The present day strength and correlation length of the magnetic fields generated by primordial black holes, for the optimal case of $Q_{\star} \alpha_{\star} =0.5$. Also shown are the constraints on this parameter space from gravitational microlensing surveys~\cite{Macho:2000nvd,EROS-2:2006ryy,Niikura:2019kqi,Oguri:2017ock}, gravitational wave observations~\cite{Kavanagh:2018ggo,LIGOScientific:2019kan,Chen:2019irf}, and from the impact of accretion~\cite{Serpico:2020ehh}. Astrophysically relevant magnetic fields ($B \gsim 10^{-20} \, {\rm G}$) could be generated by primordial black holes in the mass range of $M \sim 10^{28}-10^{36} \,{\rm g}$ without violating existing constraints. For non-extremal black holes, the strength of the resulting magnetic fields would be smaller than those shown by a factor of $Q_{\star} \alpha_{\star}/0.5$.
}
\label{fig1} 
\end{figure}

In Fig.~\ref{fig1}, we plot the strength and correlation length of the magnetic fields generated by primordial black holes, for the optimal case of $Q_{\star} \alpha_{\star} =0.5$. Also shown in this figure are the constraints on this parameter space from gravitational microlensing surveys~\cite{Macho:2000nvd,EROS-2:2006ryy,Niikura:2019kqi,Oguri:2017ock}, gravitational wave observations~\cite{Kavanagh:2018ggo,LIGOScientific:2019kan,Chen:2019irf}, and from the impact of accretion on the CMB~\cite{Serpico:2020ehh} (for reviews, see Refs.~\cite{Green:2020jor,Carr:2020gox}). From this figure, we see that astrophysically relevant magnetic fields ($B \gsim 10^{-20} \, {\rm G}$) could potentially have been generated by primordial black holes  with masses in the range of $M \sim 10^{28}-10^{36} \,{\rm g}$, without violating any existing constraints. The smallest mass in this range corresponds to the horizon mass at $\sim 100$ GeV, so this should be interpreted as an upper bound on the range of viable production temperatures. A lower bound on $T_i$ comes from the requirement that the black holes discharge prior to BBN, $T_i \gtrsim 10$ MeV. Throughout this mass range, once the black holes discharge, their Hawking radiation is negligible, so this population  constitutes a fraction of the dark matter today~\cite{Hawking:1974sw,Gibbons:1977mu}.


In this region of parameter space, the correlation length of the present day magnetic field falls in the range of $\xi \sim 10^{-6}-10^{-1} \, {\rm Mpc}$. Across this range of values, the magnetic fields are predicted to have survived the effects of magnetic dissipation and diffusion~\cite{Durrer:2013pga,Banerjee:2004df,2009PhRvD..80l3012N,Kronberg:1993vk}. More explicitly, in order to avoid early magnetic dissipation the present day field should satisfy \cite{batista2021}
\be
\xi_0 \gtrsim 10^{-7} \, {\rm Mpc}   \brac{  \langle B_0 \rangle }{ 10^{-15}\, \rm G}~.
\ee
This condition is easily fulfilled for the relevant parameter space in Fig.~\ref{fig1}. Thus, in this regime we are justified in considering only Hubble expansion in translating the early universe field to its present day value.

Though too heavy and subdominant to be of interest as dark matter candidates, and so not included in figure, there is additional viable parameter space at small $\Omega_{\rm BH}/\Omega_{\rm DM} \lesssim 10^{-5}$ and to the right of the accretion bound, which truncates at $M \sim 5 \times 10^{37} \, \rm{g}$ (see e.g. \cite{Carr:2020gox}). The existence of this additional parameter space is important nevertheless in light of updated lower bounds on the intergalactic magnetic field (IGMF)  strength based on observations of electromagnetic cascades from blazars. The non-detection of GeV-scale gamma-ray emission from electromagnetic cascades initiated from extragalatic TeV sources can be used to place lower bounds on the IGMF, as first argued by \cite{Neronov:2010}, who claimed a lower bound of $B \geq 3 \times 10^{-16}\, \rm{G}$. This bound has since been improved upon by analysis of VERITAS \cite{Veritas:2015} and \textit{Fermi}/LAT \cite{Ackermann:2018} data. Most recently, incorporating data from MAGIC \cite{MAGIC:2023} has improved the bound to $B \gtrsim 10^{-14} \, \rm{G}$. This is much more robust than previous estimates as it does away with the uncertainty in variability of the TeV band source flux. However, it still suffers from some uncertainty in the fraction of power carried away by plasma instabilities which develop from interactions between the generated particle stream and mimic the effect of the IGMF in reducing the expected cascade flux.

\section{Charged Black Holes and Chemical Potentials}
\label{sec:CP}

Thus far, we have remained agnostic regarding the origin of the Kerr-Newman black holes. We present here one concrete model capable of endowing an existing population of primordial black holes with electric charge. Of course it's very difficult to create black holes with geometrically significant charge in the early universe. In a cosmological setting, any net charge would be quickly neutralized by the surrounding plasma, which must\footnote{Black holes may violate global symmetries, but not gauge symmetries, and so any charge taken up by the black hole must be lost by the plasma.} have an opposite compensating charge to maintain the charge neutrality of the universe. Even if one were to consider a charged black hole in vacuum, its charge is expelled exponentially quickly through Hawking radiation or Schwinger pair production~\cite{Hiscock1990}. A population of charged black holes thus requires the introduction of new physics.

The Hawking radiation of electrically neutral black holes is symmetric with respect to the production of particles and anti-particles. By contrast, charged black holes preferentially radiate particles with the same sign charge as the black hole. To understand why, consider the flux spectrum for a Kerr-Newman black hole, which follows the thermal distribution~\cite{Hawking:1974sw,Page:1976df}:
\begin{equation}
	dN \sim \frac{d\omega}{\exp \left[ (\omega - m \Omega - q \Phi)/T_{\rm BH} \right] \mp 1} \,,
\end{equation}
where $\omega$ and $m$ are the energy and angular momentum of an emitted particle, $\Omega$ and $\Phi$ are the angular velocity and electrostatic potential of the black hole, and $\mp$ refers to bosons and fermions, respectively. We can identify $\mu_q \equiv q \Phi$ with a chemical potential, biasing\footnote{From this expression, we also see that $m \Omega$ acts in a similar manner, leading the black hole to preferentially expel particles whose angular momentum is aligned with the black hole. Thus, the black hole will shed both quantities as it evaporates, evolving towards a neutral, non-rotating state.} the emission of particles whose charge is aligned with the black hole. 

This chemical potential is actually sourced by the electromagnetic potential $A_\mu$ of the Kerr-Newman black hole itself. A particle of charge $q$ at the horizon couples to $A_\mu$ as $\mathcal{L} \supset - q A_\mu J_{\rm EM}^\mu$.  Since the time-like component couples to the charge density $J_{\rm EM}^0$, we can identify the combination $- q A_0 |_{r_+}$ with a chemical potential, $\mu_q$:
\begin{equation}
\label{approach1}
	\mathcal{L} \supset - q A_0 J_{\rm EM}^0 \equiv \mu_q J_{\rm EM}^0 \,.
\end{equation}
Using Eqs.~(\ref{apotential}) and (\ref{EMpotential}), one can verify that this combination $\mu_q = - q A_0 |_{r_+}$ is identical to $\mu_q = q \Phi$, and so we see this is self-consistent. 

Just as the intrinsic chemical potential of the Kerr-Newman black hole allows it to shed its charge, one can imagine charging up a black hole (or maintaining a black hole in a charged state) by means of an external chemical potential. If such a chemical potential is greater than that of the black hole itself, then the black hole will build up charge until it reaches an extremal state. 

One possible mechanism for realizing such a chemical potential involves a new scalar field $\phi$ derivatively coupled to the electromagnetic current via:
\begin{equation}
\label{deriv}
	\mathcal{L} \supset  \frac{1}{\Lambda} \partial_\mu \phi J_{\rm EM}^\mu  \,.
\end{equation}
Such an operator generically arises in the effective Lagrangian description as the most relevant coupling to the Standard Model if $\phi$ has an approximate shift symmetry. In this case $\Lambda$ would be interpreted as the symmetry breaking scale. If $\phi$ is initially displaced from the origin and begins rolling in the early universe, its time derivative will source an effective chemical potential for charged particles, $\mu_\phi \equiv \dot{\phi}$, leading the black hole to preferentially absorb particles with charge of a particular sign. Magnetic field generation will occur during the period in which the external chemical potential is active because the scalar field is rolling. Once $\phi$ stops rolling at temperature $T_i$, the chemical potential will vanish and the black hole will quickly expel its charge, thereby returning to a neutral state.


\section{Summary and Conclusions} 
\label{sec:summary}
  
In this article, we have studied the possibility that cosmological magnetic fields may have been generated in the early universe by a population of primordial Kerr-Newman black holes. We find that black holes near extremality ($\alpha_\star Q_\star \sim 0.5$) in the mass range of $M \sim 10^{28}-10^{36} \,{\rm g}$ would have been capable of producing present day magnetic fields that are as large as $B \sim 10^{-15} \, {\rm G}$. Black holes at larger masses $M \gtrsim10^{38} \, \rm{g}$ and smaller fractional abundance could have produced even larger fields $B \gtrsim 10^{-14} \, {\rm G}$. The corresponding correlation lengths are sufficiently large that these fields would have survived the effects of early magnetic dissipation and diffusion. Thus these fields could have seeded larger galactic and intergalactic fields through the dynamo mechanism.
  
In order to generate a magnetic field, the black holes in this scenario must be both spinning and electrically charged. Throughout most of our analysis, we have remained agnostic as to the origin of these Kerr-Newman black holes. While it is straightforward to create spinning black holes through the mergers of an initial population of Schwarzschild black holes~\cite{Hooper:2020evu}, it is more challenging to explain how these black holes acquire an appreciable net electric charge in the early universe. As discussed in Sec.~\ref{sec:CP}, one possibility involves a rolling scalar field which dynamically generates a chemical potential for charged particle species, thereby biasing the charge distribution of Hawking radiation and the net flow of charge into the black holes. We leave the model building that concretely realizes such a scenario for future work.

Our analysis has also largely omitted the complicated magnetohydrodynamics (MHD) processes that govern the evolution of magnetic fields in a hot thermal plasma. While a proper treatment depends on the spectral shape and is beyond the scope of this letter, we note that large scale MHD decay will generically serve to increase the correlation length and power on large scales~\cite{Son:1999}. This occurs because power on small scales dissipates more efficiently and because there exists a weak inverse cascade which shifts energy away from the dissipation scale, even in the case of non-helical freely decaying MHD turbulence~\cite{Vachaspati:2021}. Consequently, our estimates for the correlation length, which took into account only the passive expansion of the universe, are conservative, and we can all the more safely neglect the effects of diffusion and dissipation. Note though that there may still be phenomena such as turbulence which could serve to tangle magnetic field lines and excite dissipative modes. Turbulent decay will generically serve to reduce the total magnetic energy density~\cite{Brandenburg:2016}, such that the figures for the field strength in the previous section should be seen as an upper estimate.

Another feature omitted from our treatment is the helicity of the magnetic field, $\mathcal{H} = \int d^3x \, \vec{A} \cdot \vec{B}$. Magnetic helicity is conserved during evolution and can result in interesting effects. One can see from the form of the vector potential in Eq. (\ref{EMpotential}) and magnetic field in Eq. (\ref{eq:fields}) that there is no intrinsic magnetic helicity associated with the Kerr-Newman solution. If the initial velocity field has a non-vanishing kinetic helicity, however, it is possible for the magnetic field on large scales to develop a corresponding helicity, with an opposite helicity developing on small scales by conservation of helicity~\cite{Brandenburg:2017}. A more thorough investigation of these effects, including numerical simulations, is left to future investigation.

Finally, we note that this scenario predicts a nontrivial relationship between the primordial magnetic field parameters and the merger rate for the progenitor black hole population. This suggests that the primary observable signature of the model would be a correlation between the parameters of the cosmological magnetic field and the stochastic gravitational wave background coming from the merger of primordial black hole binaries. Presuming a monochromatic mass function, the present day merger rate of binaries which formed in the early universe can be roughly estimated as \cite{Raidal:2017}:
\begin{equation}
	R_3(t_0) \sim 80 \left( \frac{\Omega_{\rm BH}}{10^{-3}} \right)^{53/37} \left( \frac{M_{\rm{BH}}}{10^{34} \, \rm{g}} \right)^{-32/37} \, \rm {Gpc}^{-3} \rm{yr}^{-1} \,,
\end{equation}
where we have conservatively set the local density contrast at decoupling $\delta_{\rm dc}$, which quantifies the effect of clustering, to $\sim 1$. This can then be compared with the rate indicated by LIGO observations, $12-213\, \rm{Gpc}^{-3} \rm{yr}^{-1}$ \cite{LIGO:2017}. A more thorough analysis depends sensitively on the clustering dynamics and is left to future investigation.

 \bigskip

\begin{acknowledgments}  

DH and GK are supported by the Fermi Research Alliance, LLC under Contract No.~DE-AC02-07CH11359 with the U.S. Department of Energy, Office of Science, Office of High Energy Physics. AI is supported by the U.S. Department of Energy, Office of Science, Office of Workforce Development for Teachers and Scientists, Office of Science Graduate Student Research (SCGSR) program, which is administered by the Oak Ridge Institute for Science and Education (ORISE) for the DOE. ORISE is managed by ORAU under contract number DE‐SC0014664.

\end{acknowledgments}

\bibliography{PBH_Magnetogenesis}

\end{document}